\documentstyle[DL99, times, psfig,graphicx]{article}
\newcounter{example}
\newcounter{query}
\newcounter{property}
\newcounter{definition}
 
\newcounter{theorem}
\newcounter{lemma}
\newcounter{corollary}
\newcounter{assumption}

\newtheorem{definition}{Definition}
\newtheorem{example}{Example}

\def\VNP.#1.#2.#3.#4{Vol.~#1, No.~#2, pp.~#3--#4}
\def\NP.#1.#2.#3{No. #1, pp. #2--#3}
\def\VP.#1.#2.#3{Vol. #1, pp. #2--#3}
\def\NA.#1.#2{#1{\footnotesize #2}}
\def\LEQ.#1.#2.#3{#2\!\leqslant\!#1\!\leqslant\!#3}
\def\OP.#1.#2.#3{#1\!#2\!#3}

\def\VLDB.#1{{\em Proceedings of the #1 International Conference on Very 
	Large Data Bases\/}}

\def\EndOfProof{\nolinebreak\ \hfill\rule{1.5mm}{2.7mm}}

\begin{document}

\newpage
\sloppy

%-----------------------------------------------------------------------
\title{{\LARGE\bf ZBroker : A Query Routing Broker for Z39.50 Databases}}
\author{Yong Lin\hspace{5ex} Jian Xu\hspace{5ex}
	Ee-Peng Lim  \hspace{5ex}  Wee-Keong Ng\\
	Centre for Advanced Information Systems\\
	School of Applied Science, Nanyang Technological University\\
	Nanyang Avenue, Singapore 639798\\
	Email: \{P146046183, P140977971, aseplim, awkng\}@ntu.edu.sg
	\\[1ex]}

\maketitle

\abstract
A query routing broker is a software agent that determines from a large
set of accessing information sources the ones most relevant to a user's
information need. As the number of information sources on the Internet
increases dramatically, future users will have to rely on query routing
brokers to decide a small number of information sources to query without
incurring too much query processing overheads.  
In this paper, we describe a query
routing broker known as ZBroker developed for bibliographic database
servers that support the Z39.50 protocol.  ZBroker samples the content of
each bibliographic database by using training queries and their results,
and summarizes the bibliographic database content into a knowledge base.
We present the design and implementation of ZBroker and describe
its Web-based user interface.

\paragraph{KEYWORDS:} query routing, bibliographic databases, digital libraries

\section{1. INTRODUCTION}

As the number of information sources increases rapidly on the Internet,
users are beginning to experience difficulties locating the relevant
information sources to meet their search requirement. These information
sources could be document collections, SQL databases, or other kinds of
databases.
Although many web search engines are available on the Internet, most
of them are only useful for discovering individual web pages instead
of information sources such as document collections and SQL databases.
Hence, these search engines cannot be easily extended to
index the content of information sources.

Given a user query and a set of information sources at different
locations, {\em query routing} refers to the general problem of
selecting from a large set of accessible information sources the ones
relevant to a given query, and evaluating the query on the selected
sources.  The software agent that performs query routing on the
Internet is known as a {\em query routing broker}.
In this paper, we describe a query routing broker developed for
bibliographic databases supporting {\em Z39.50 protocol}.
This query routing broker, called {\em ZBroker}, is currently
being developed at the Centre for Advanced Information Systems,
Nanyang Technological University.

Z39.50 is an application-level communication protocol\cite{z3950}
adopted by the International Organization for Standardization (ISO) and is
designed to support remote access to bibliographic databases maintained
by public libraries.  Z39.50 is widely supported by library system vendors.
It specifies a uniform interface for a Z39.50 client application to
search and retrieve information from bibliographic databases managed
by different Z39.50 servers.  In other words, Z39.50 allows 
bibliographic databases implemented by different vendors on different
computer hardware to look and behave the same to the client application.
A bibliographic database with Z39.50 query support is also known as
a {\em Z39.50 database}.
A listing of public libraries supporting Z39.50 access can be found
at {\em http://www.mun.ca/library/cat/z3950hosts.htm}.

The objective of the ZBroker project is to design and implement a
software agent capable of routing bibliographic queries on the Internet
populated by hundreds of bibliographic databases.  
A bibliographic database provides important meta-information about the 
material found in a library.  Apart from allowing users to physically locate
reading materials on the shelves, bibliographic databases play
an important part in the learning and research processes that users
have to go through.  Usually, a user first refers to a bibliographic
database before identifying the reading material that is relevant to
his or her interest.  In the context of Internet, a researcher can 
search multiple bibliographic databases to reveal reading
material that may not be available in his or her library.
The researcher can further request inter-library loan to obtain
the desired material.  

\subsection {Scope of Work}

To enable Internet users to easily perform query routing tasks
for bibliographic databases, a few functionalities have to be
supported by the ZBroker:
\begin{itemize}
  \item {\em Acquisition of content knowledge}:
	This refers to summarizing the content of a set of 
	bibliographic databases in order to capture the knowledge 
	about their content.
  \item {\em Database ranking based on a given bibliographic query}:
	When a bibliographic query is given to ZBroker, it should
	rank the bibliographic databases according to their
	relevance computed from the respective content knowledge.
	At present, the degree of relevance is measured by the
	estimated result size returned by a bibliographic database
	for the given query.   
  \item {\em Maintenance of content knowledge}:
	As the bibliographic databases evolve, their content
	may change and this leads to the need to maintain the
	content knowledge previously acquired by the ZBroker. 
  \item {\em User-friendly interface}:
	ZBroker should provide an easy-to-use web interface
	to the users.  The user interface can support 
	queries on multiple Z39.50 databases selected by the
	users who will be guided by the database ranks 
	computed by ZBroker. 
\end{itemize}

Although Z39.50 supports queries on multiple bibliographic attributes
such as title, author, subject, ISBN, ISSN, etc., not all 
Z39.50 servers implement the same query capabilities.
In our project scope, we first consider boolean queries that involve
title, author and/or subject attributes.  These are the attributes
that are searchable in most Z39.50 databases. 
The query results returned by the Z39.50 databases are unranked.
Once a set of Z39.50 databases have been ranked for a given query,
ZBroker will present the location and rank information of the 
respective Z39.50 databases.
When a bibliographic database fails to support a particular
query, it will be assigned the lowest (least relevant) rank by
ZBroker.

In this project, we experimented the query sampling technique to acquire
the content knowledge of a Z39.50 database.  The sampling is
performed by submitting training queries to the Z39.50 database
and summarizing the returned results.  The salient feature of
this sampling technique is that it does not require modification
to the existing Z39.50 servers, thus not compromising the 
autonomy of databases owned by different institutions.  
Nevertheless, it is crucial that the sampled content knowledge 
demonstrates a certain level of accuracy in order for ZBroker 
to perform well.

\subsection {Paper Outline}

The rest of this paper is structured as follows.  
In Section 2, we provide a brief survey of the related work.  
Section 3 presents the system architecture of ZBroker. 
Section 4 describes the information captured by the ZBroker's
content knowledge and database ranking technique.
The sampling technique and content knowledge maintenance are 
given in Sections 5 and 6 respectively.
Section 7 gives an overview of the ZBroker's user interface.
The performance issues of ZBroker are discussed in Section 8. 
Section 9 concludes the paper and presents the future work.

\section {2. RELATED WORKS}

Query routing in the context of document collections has been 
studied by several researchers\cite{gloss:94,cori:95,voorhees:95,dik:97}.
Most of these works involve the performance evaluation of different
query routing techniques. 
Nevertheless, to our best knowledge, there has not been a query 
routing broker developed for Z39.50 databases.
In this section, we survey some system research works that
are related to query routing.

\subsection {Federated Searcher}

{\em Federated Searcher} is a query routing broker implemented by 
Virginia Tech for mediating user queries to multiple heterogeneous 
search engines\cite{fox:98}. Each search engine has to provide a description 
about its site using a specially designed XML markup language
known as SearchDB.  Federated Searcher has been implemented for 
the Networked Digital Library of Theses and Dissertations 
(NDLTD) (see {\em http://www.theses.org}). Instead of storing 
the content information about each site, Federated Searcher 
captures the types of documents indexed by each site, 
interface information, and general information about the search 
engine used by each site.  Federated searcher also includes a 
translation request protocol that facilitates multilingual searches.
Note that the site description required by Federated Searchers
is manually created.  Hence, it is difficult for the site description
to capture complete content knowledge about the site.

\subsection {STARTS Protocol}

{\em STARTS Protocol} is proposed by Stanford University researchers to
support meta-searching or query routing on the Internet\cite{starts:97}.
The protocol defines source selection, distributed query evaluation 
and query result merging as the three main query routing tasks.  
In order to handle largely incompatible search engines, 
STARTS requires a standard set of metadata to be exported from different 
database servers. The acquisition of these metadata information
has been incorporated into STARTS.  When a database server supports 
STARTS protocol, it is able to provide detailed statistics about
its collection upon request by a query routing broker, thus allowing
the broker to have first hand detailed information about the 
database content.  Nevertheless, it may not be possible to have
all database servers supporting STARTS and the exchange of detailed
site information as it may require modification to the existing
database servers and their existing information retrieval protocols.

While STARTS requires some level of cooperation among database
vendors and owners in order to have their database servers
providing useful meta-data for query routing purposes, ZBroker
adopts a less intrusive approach to solicit meta-data from the
database servers by sending queries to probe the database content.
We believe, however, that the two approaches can co-exist together
to provide a more complete set of query routing solutions for 
different types of database servers.  

\subsection {Glossary of Servers Server Project}

In the GlOSS ({\em Glossary of Servers Server}\/) project\cite{gloss:99,
broker.stru:96,gloss:94}, a keyword-based distributed database broker
system is proposed and it can route queries containing a set of {\em
keyword field-designation} pairs where the {\em field-designation} could
be {\em author}, {\em title}, etc. The number of documents containing
each term for each {\em field-designation} is stored and used to
estimate the rank of each database. The databases considered by GlOSS
are unstructured document collections. The main assumption behind GlOSS
is that {\em terms}\/ appearing in different documents of a collection
follow independent and uniform distributions. 
GlOSS has been implemented for the Networked Computer Science 
Technical Reference Library
(see {\em http://www.ncstrl.org} for the NCSTRL project).

\section{3. SYSTEM ARCHITECTURE}
\label{sect:arch}

\begin{figure}[htb]
\centering
\includegraphics[width=8cm]{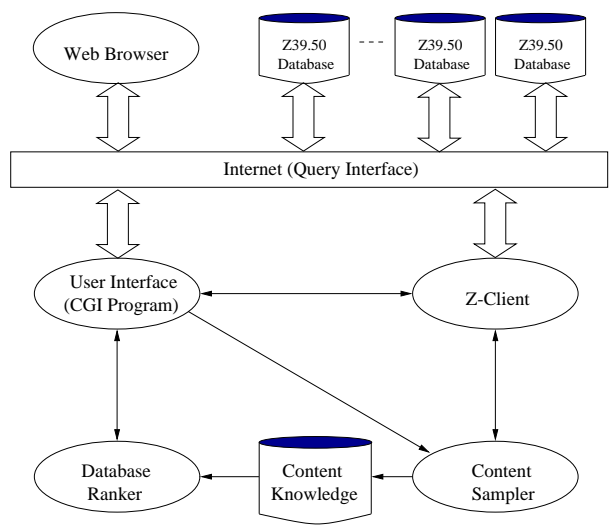}
\caption{\sf ZBroker System architecture}
\label{fig:architecture}
\end{figure}

In this section, we present an overall design of ZBroker.
ZBroker consists of a few core modules as shown in 
Figure\ref{fig:architecture}.  

The ZBroker's {\em web-based user interface}
is driven by a Common Gateway Interface (CGI) Program which
allows Internet users to input bibliographic queries to be
routed.  The module also presents query results returned from
bibliographic databases selected by the users.
A detailed description of ZBroker's user interface
will be given in Section 7.

%--- ZClient
To access remote bibliographic databases using the Z39.50 protocol,
a common gateway supporting Z39.50 application programming interface
(API) is required.  In the ZBroker project, we have chosen to use
{\em Z-Client}, a gateway implemented by Harold Finkbeiner\footnote{
The software is available at 
{\em ftp://lindy.stanford.edu/pub/z3950/zclient.tar.gz.}}.  
Built upon Z39.50 API functions, Z-Client provides methods that 
submit queries to specified Z39.50 databases, and parse the 
returned query results into some required formats.  
Each Z39.50 database is identified by its Internet address, 
port number, and a database id.

%--- Database Ranker
{\em Database Ranker(DB-Ranker)} is designed to rank Z39.50 databases 
according to their relevance to the user query submitted through
the user interface and the CGI program.
The database ranks are computed based on statistical information
about the content of participating Z39.50 databases.  
These information are collectively known as the {\em Content Knowledge}.
The computed database ranks will be returned to the user interface program
for meaningful presentation.  The database ranking formula
adopted by DB-Ranker resembles that of the GlOSS but it only
makes use of smaller set of database records sampled by training
queries.  A detailed description of this ranking formula will be
given in Section 4.
 
%--- Content Sampler
{\em Content Sampler} samples the participating Z39.50 databases 
and construct the content knowledge about them.  The sampling
procedure consists of submitting training queries to the Z39.50
databases via Z-Client and collecting the query results.
To cope with the evolving content of these Z39.50 databases and
to enlarge the range of sampled bibliographic records,
Content Sampler accumulates user queries submitted to ZBroker
and selects the appropriate ones as new training queries.

\begin{figure}[htb]
\centering
\includegraphics[width=8cm]{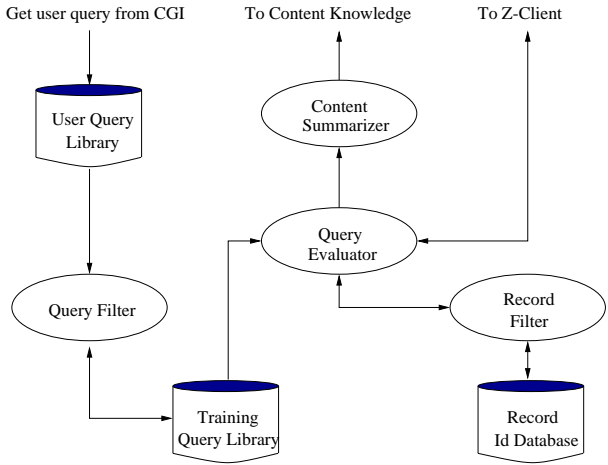}
\caption{\sf Architecture of Content Sampler}
\label{fig:sampler}
\end{figure}

Figure~\ref{fig:sampler} illustrates the four sub-components 
of Content Sampler, namely, {\em query evaluator}, {\em content
summarizer}, {\em record filter}, and {\em query filter}.
The query evaluator is responsible for retrieving training
queries from a training query library, formatting and submitting 
them to the remote Z39.50 databases.  This is done via 
Z-Client.  Upon receiving a training
query result set, the query evaluator passes the bibliographic
records in the result set to the content summarizer which updates
the content knowledge about the database concerned.
To ensure that only unique bibliographic records are given to
the content summarizer, the record filter will be invoked to
discard bibliographic records that have already been returned
by previous training queries.  We maintain for each Z39.50
database a {\em record id database} which keeps the system ids
of all records that have been examined by the content summarizer.

In Figure~\ref{fig:sampler}, we also indicate that user queries
submitted to the ZBroker user interface can be captured by the
{\em User Query Library}.  The user queries accumulated in this
manner can be later deployed as training queries if the user
queries can potentially return new records not yet considered by
the content summarizer.  While the records can be filtered by the
record filter, it would save the content sampler much processing
overheads if the filtering can be performed on the user queries
before they are added to the training query library.
This filtering task is the responsibility of the query 
filter.  We will describe query filtering in more detail in 
Section 5.

\section{4. CONTENT SUMMARIZATION AND DATABASE RANKING}
\label{sect:rank}

In this section, we describe the ranking technique adopted by
the ZBroker's database ranker.  The ranking technique in turns
determines the summarized information to be included by the content
knowledge.

In ZBroker, databases are ranked according to their estimated
relevance scores computed from some statistics about the databases.
The relevance score formula, known as {\em Training Query Result
Summary using GlOSS (TQG)}, resembles that proposed by the GlOSS
project at Stanford University except that it is applied to a
set of records sampled from the database.  
Given a query $q$, the TQG formula assigns the estimated relevance
score, ${\mathcal E}_{db_i,q}$, to database $db_i$ as follows:

\begin{definition}
\label{def:tqg}
\[{\mathcal E}_{db_i,q} = \widehat{Size}_{(db_i,q)} = N'_i
\prod_{k=1}^{|A|}
		      \prod_{j=1}^{|A_k|}
  		 \frac{TF'_{i,j,k}} {N'_i}\]
where $|A_k|$ denotes the number of search terms on attribute $A_k$ 
specified by the query $q$; $|A|$ denotes the number of attributes
involved in the query $q$; $TF'_{i,j,k}$ denotes the tuple
frequency of the $j$th search term in attribute $A_k$ 
computed from the set of records sampled from database $db_i$; and 
$N'_i$ denotes the total number of sampled records for database 
$db_i$.\EndOfProof
\end{definition}

Note that the tuple frequency of a search term in an attribute
refers to the number of records in the database containing the
term in the specified attribute.
Clearly, the number of sampled records $N'_i$ is always smaller
than $N_i$, the actual number of records in database $db_i$.
Since $TF'_{i,j,k} \leq N'_i$, the estimated relevance score 
ranges between 0 and $N'_i$.  

Therefore, both $TF'_{i,j,k}$'s and $N'_i$ have to be computed by
the content summarizer for each database and be kept in the 
content knowledge for ranking purposes.
In the following section, we will describe how database content
sampling is performed by the query evaluator and record filter
of ZBroker.

\section {5. DATABASE CONTENT SAMPLING AND RECORD FILTERING}
\label{sect:sample}

To sample the content of databases, training queries has to be 
generated and be used to extract records from the databases. 
There are essentially two possible approaches to create such
training queries.  One can either create synthetic training queries 
or collect user queries as training queries.
In the current implementation, we have chosen to sample database content
using both kinds of queries.  
Synthetic training queries have been generated to {\em bootstrap}
the construction of content knowledge.  User queries, on the other
hand, have been used together with the initial set of synthetic
queries to capture changes in the Z39.50 databases' content as
they evolve.

In this section, we describe the generation of synthetic queries,
and the filtering processes that are applied to the user queries and
the bibliographic records returned by these queries.

%---------
\subsection {Synthetic Training Query Generation}

In order to ensure that the synthetic queries provide a good
coverage of the Z39.50 database content, we require them
to be simple enough such that a fair number of records can
be sampled from the particpating databases.
Hence, small number of search terms are used in these synthetic queries.
At present, ZBroker keeps a database of 3000 synthetic training queries
generated from the bibliographic records of the Nanyang Technological
University (NTU) Library\footnote{The web page of Nanyang Technological
University library is available at:(http://web.ntu.ac.sg/library/).}.
We downloaded 218,000 bibliographic records from the NTU library, and
generated the synthetic queries by repeating the following steps:

\begin{itemize}

  \item {\em Step 1:}\/ Randomly select a bibliographic record.

  \item {\em Step 2:}\/ Randomly decide whether to use {\em title}\/,
  {\em subject}\/ or both attributes.

  \item {\em Step 3:}\/ Extract from the selected record up to four
  distinct terms from the chosen attribute(s).  Since a bibliographic
  record may have several values for the subject attribute, all
  extracted subject terms must come from the same value.
  Moreover, stop words are not allowed to be included among the
  extracted terms.

  \item {\em Step 4:}\/ The extracted title and/or subject terms
  form the search terms for a new synthetic training query.

\end{itemize}

\subsection{Deriving New Training Queries from User Queries}

When a user query can potentially return new records not yet considered
by content sampler, Z39.50 will deploy the user query as a new
training query and use it together with other training
queries in sampling the database content.
Before a user query can be deployed as a training
query, we have to determine if it returns a result set
that is a subset of that of any other existing training query.

\begin{definition}
A query $q_1$ is said to be {\bf result-subsumed} by query $q_2$ with
respect to a database DB if the result set of query $q_1$ from DB is
a subset of those of query $q_2$.\EndOfProof
\label{def:subsumed}
\end{definition}

Owing to the heterogeneity of query formats and database content,
it is difficult to decide if a user query is result-subsumed by some
existing training query.  
In our ZBroker project, we therefore adopt a more restrictive
definition of subsumption relationship between queries.
This restrictive definition is possible as each user
query handled by ZBroker consists of conjunction of
search terms and are known as {\em conjunctive queries.}

\begin{definition}
Two selection predicates $p_1 \equiv (A_1 = val_1)$ and $p_2 \equiv (A_2
= val_2)$ are said to match ({\bf predicate matching}) 
if $A_1 \equiv A_2$, where $A_1$ and $A_2$
are attribute names, and each $val_i$ ($i=1$ or $2$) represents a conjunctive
set of terms.\EndOfProof
\end{definition}

\begin{definition}
Let $q$ be a conjunctive query, $P(q)$ denotes all search predicates
in a query $q$.  A query $q_1$ is {\em predicate-subsumed}
by another query $q_2$ if $P(q_2) \subset P(q_1)$
(i.e., every predicate $p_{2j}$ in $q_2$ has
a matching predicate $p_{1j}$ in $q_1$) and
$val_{2j} \subset val_{1j}$,
where $val_{ij}$ refer to the distinct term set in predicate $p_{ij}$
for query $q_i$ ($i=1$ or $2$).\EndOfProof
\end{definition}

\begin{example}
\em
Let query $q_1$ = (title = ``digital library'') and query $q_2$ = (title
= ``digital''), $q_1$ is predicate-subsumed by $q_2$ as the term
``digital'' in $q_2$ can also be found in the matching predicate in
$q_1$.\\
Query $q_1$ = (title = ``database management project'' and subject =
``database management'') is predicate-subsumed by $q_2$ =
(title = ``database management'').\\
On the other hand, $q_3$ = (title = ``computer management'') is not
predicate-subsumed by $q_4$ = (title =``bussiness management'').\EndOfProof
\label{exam:qrule}
\end{example}

Hence, when a user query is found not predicate-subsumed by
any existing training query, it will be included as a new
training query.  The verification of predicate-subsumption is
performed by the query filter as shown in Figure~\ref{fig:sampler}.
% (*** What happen why an existing synthetic query is predicate-
% subsumed by a user query?)
The query filter examines all user queries that have been
logged in the user query library,
and determines the predicate-subsumption relationship
between the existing training queries and these queries.

\subsection{Record Filtering}

Despite the effort in ensuring that no training query is
redundant, it is still possible for two training queries
to return overlapping result sets.  Such duplicate bibliographic
records, if not discarded, will lead to incorrect tuple frequencies
in the content knowledge.

In ZBroker, a {\em Record Id Database} is therefore
maintained for each Z39.50 database and it is used by the
{\em record filter} module to detect and discard duplicate
duplicate records.
For most Z39.50 databases, unique ids are assigned
to their records and it is relatively easy to extract these
ids from the training query results and keep them
in the record id databases for record filtering.
Nevertheless, not all Z39.50 databases provide such record ids.
For a Z39.50 database that does not provide record ids,
ZBroker determines the uniqueness of records by using
either both author and title, or ISBN.
To efficiently filter records using the record id databases,
the record ids are stored as binary search trees in the
databases.  Only for those records that does not come
with ids, binary search trees for author, title and ISBN
are created. 

\subsection{Global Term Id Assignment}

As mentioned in Section 4, the ZBroker's content 
knowledge stores for each Z39.50 database the tuple frequencies 
for terms appearing in different attributes of the database.
To efficiently access the tuple frequencies, we require ids to
be assigned to the terms.  If a different set of term ids are
created for each Z39.50 database, one would require a separate
disk space for storing the mappings between terms and their ids.
To reduce the amount of storage space required, 
ZBroker assigns unique {\em global term ids} to the
terms found in all Z39.50 databases.  Hence a {\em Global Term 
Dictionary} has been created to maintain the term ids for each 
attribute. 

Using the global term dictionary of an attribute, a single unique 
global term id can be assigned to a term found in the attribute
of any Z39.50 databases.
A {\em Global Term Id Manager}, which is part of content sampler,
is responsible for assigning 
the global term ids.  Using the global term ids, the tuple
frequencies in the content knowledge can be retrieved efficiently.

When ZBroker constructs the content knowledge, several
instances of content samplers may run simultaneously, one for each
Z39.50 database.  To regulate concurrent accesses to the global 
term dictionary, each content sampler process has to acquire a lock 
before the dictionary can be accessed.  By holding the lock, it
prevents other content samplers from accessing the dictionary
thereby protecting the integrity of content knowledge.

\section{6. CONTENT KNOWLEDGE MAINTENANCE}

The maintenance of content knowledge is essential for two reasons.
It allows ZBroker to keep an up-to-date content knowledge about
the participating Z39.50 databases.  It also improves the 
accuracy of database ranking as new training queries are submitted.
In the operational mode, ZBroker receives new incoming queries submitted
by the users.  The routing of these user queries has to be performed
based on the content knowledge constructed prior to these queries.
Since not all records that are relevant to the queries have been 
captured by the content knowledge, it is possible that inaccurate
database ranks will be suggested by ZBroker. 
To maintain the content knowledge, ZBroker adopts two updating
schedules for the content knowledge, i.e., daily and monthly updates.

\begin{itemize}
	\item {\em Daily Update}\/ 
	To improve the the accuracy of database ranking,
	ZBroker collects further statistics about each Z39.50 database
	by submitting a batch of filtered user queries to all 
	Z39.50 databases at the end of each day.  
	The new information collected enables ZBroker to 
	return more accurate database ranking when similar user
	queries are given in the near future. 

	\item {\em Monthly Update}\/ 
	When new records are added to the Z39.50 databases,
	it is not possible for ZBroker to be alerted.
	Hence, ZBroker must re-sample the Z39.50 databases using
	all training queries, including the original
	synthetic training queries and filtered user queries.
	At present, ZBroker performs this update monthly.
\end{itemize}

\section {7. WEB-BASED USER INTERFACE}
\label{sect:gui}

In this section, we describe the web-based user interface 
of ZBroker that has been implemented.
The user interface is now available at 
{\em http://www.cais.ntu.edu.sg:8000/$\sim$jxu/z3950}.
The main functionalities of the user interface include:

\begin{itemize}
  \item Allowing users to formulate queries to be routed
        for a list of accessible Z39.50 databases;
  \item Ranking the databases for a given user query,
        and allowing the user to select the target Z39.50
        databases for query submission. 
  \item Forwarding the user query to the selected databases
        and presenting the query results to the user.	
\end{itemize}

\begin{figure*}
\centerline{\psfig{figure=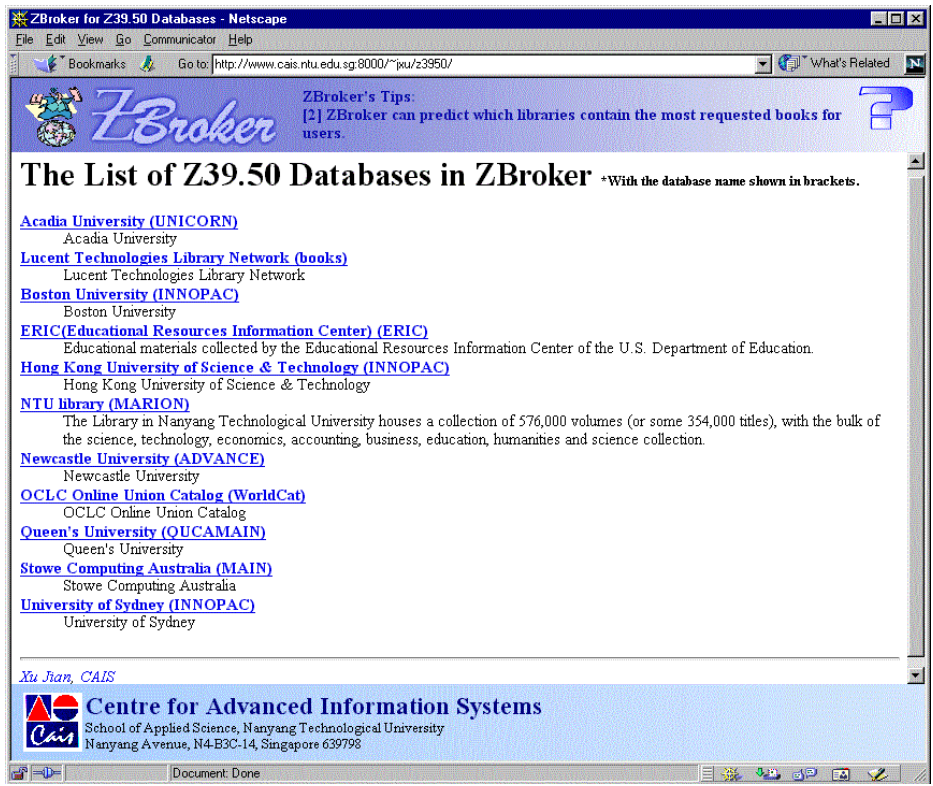}}
\caption{\sf The List of Participating Z39.50 Databases}
\label{fig:s_list}
\end{figure*}

At present, a list of eleven Z39.50 databases 
at different locations can be routed by ZBroker as shown in 
Figure~\ref{fig:s_list}. 

\subsection{Query Formulation}

\begin{figure*}
\centerline{\psfig{figure=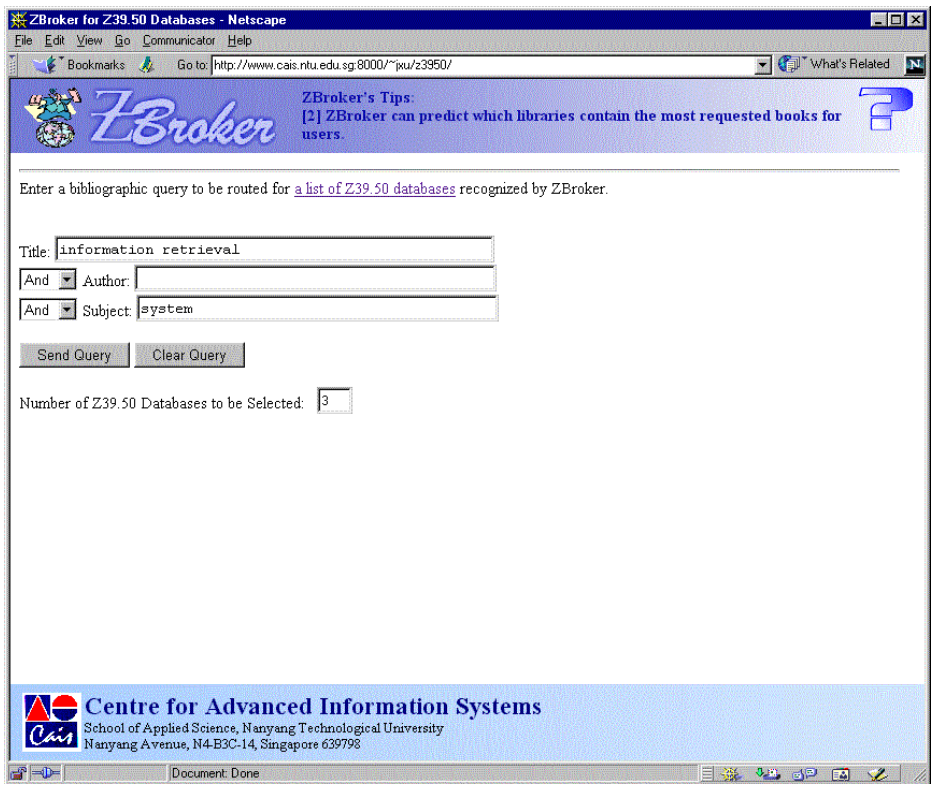}}
\caption{\sf Query Formulation Page}
\label{fig:s_query}
\end{figure*}

The user interface supports the formulation of user queries 
by requiring the users to supply search terms to the three
queryable attributes: Author, Title and Subject.  Multiple
terms supplied to a queryable attribute are combined together
with the ``and'' semantics, and the search predicates on 
different attributes are ``and'' together as well. 
Figure~\ref{fig:s_query} depicts a formulated query that 
searches for records containing ``information'' and ``retrieval''
in their titles, and ``system'' in their subject.

\begin{figure*}
\centerline{\psfig{figure=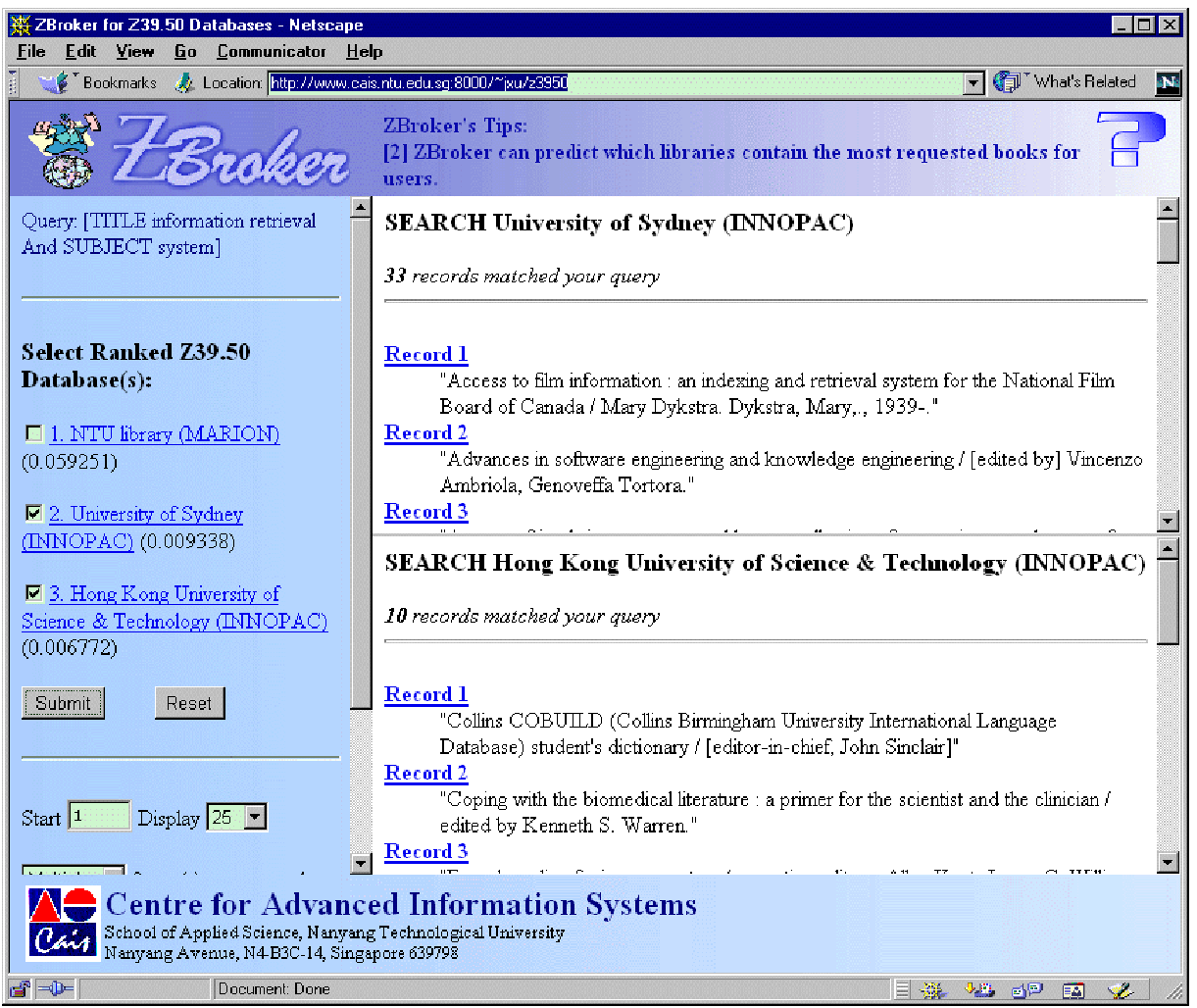}}
\caption{\sf Database Ranking and Query Result Presentation}
\label{fig:s_rank}
\end{figure*}

Once a user submits a query, the contents of the three fields
will be transferred to the ZBroker's web user-interface program.  
The CGI program extracts the query and invoke database ranker 
which in turns returns the database ids according to their ranks.
Based on the ordered database list returned by database ranker, 
the user-interface program presents the ranked databases within
the left frame of the web page as shown in Figure~\ref{fig:s_rank}. 

\subsection{Presentation of Query Results}

\begin{figure*}
\centerline{\psfig{figure=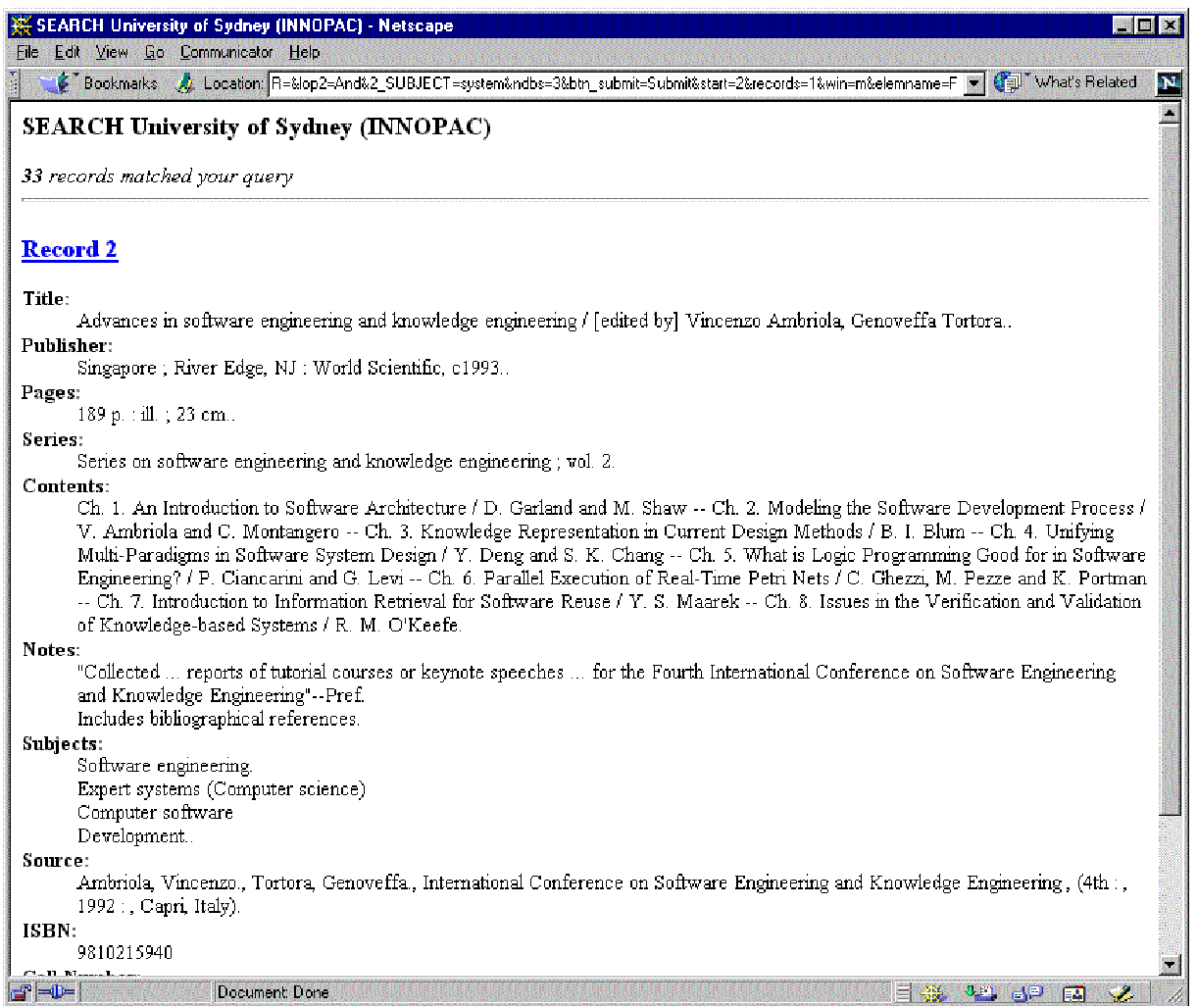}}
\caption{\sf Bibliographic Record Information}
\label{fig:s_record}
\end{figure*}

As shown in Figure~\ref{fig:s_rank}, the databases ranked by
ZBroker are listed according to their relevance scores.
At this point, the ranks are suggested by ZBroker without
sending the query to the remote Z39.50 databases.
The user can choose to submit the query to one or more
Z39.50 databases by checking the boxes for the databases,
and specifying the number of records to be retrieved from
each database.
ZBroker will forward the query to the selected databases
via Z-Client, and display the returned result sets from the 
databases in multiple frames on the right, one for each selected
database, as shown in Figure~\ref{fig:s_rank}.
User can also view the detailed information of a bibliographic
record in the result sets by following the record links
embedded in the result frames as shown in Figure~\ref{fig:s_record}.
 
\section{8. PERFORMANCE ISSUES OF ZBROKER - A PRELIMINARY ANALYSIS}

In this section, we present a preliminary analysis of the
performance of ZBroker.  The analysis focuses on examining 
the accuracy of query routing, the overheads of content sampling 
and the storage requirement of ZBroker.  
Since ZBroker can only construct
its knowledge about each Z39.50 database by drawing records from
the database using large number of training queries, it is expected
that the sampling process incurs significant amount of overheads, including
the network and CPU overheads.  In addition, ZBroker
requires storage space for its content knowledge, training queries,
global term dictionaries, and record id databases.
We will give some statistics in this section to show the amount
of sampling overheads, and storage requirement. 

\subsection{Accuracy of Database Ranking}

At present, the accuracy of database ranks suggested by ZBroker
has been encouraging.  Most of the time, ZBroker is able to
suggest relevant databases in response to a user query.
Nevertheless, since the same boolean
query can be interpreted and evaluated differently by the
particpating Z39.50 databases, one may find the actual result
sizes returned by the databases not consistent with the computed 
database ranks.    
For example, for a query involving search terms on title attribute,
some Z39.50 database servers may impose the search criteria on not
only the title attribute but also other related attributes leading to more
records included in the query results.
In these cases, a lowly ranked database may appear to return
much more records than a highly ranked database but many
of these records do not carry the search terms in their title attribute.

\subsection{Overhead of Content Sampling}

The overhead of sampling some Z39.50 databases is tabulated in 
the second column of Table~\ref{table:cost}.  
The table indicates the number of hours taken to sample each database.
The sampling overhead ranges from 6 to 159 hours (about 7 days) 
depending mostly on the network delays involved between ZBroker 
and the remote Z39.50 databases.

\begin{table*}[htp]
\footnotesize
\centering
\begin{tabular}{|c|c|c|c|}
\cline{1-4}
Database Alias & Time(hours) taken & \# of Records Sampled & Record ids storage(MBytes)\\
\cline{1-4}
acadia & 25 & 214303 &5.068\\
bell & 6 & 4432 & 0.112\\ 
boston & 59 & 592379 & 14.097\\
eric & 12 & 4935 & 0.182\\
hongkong & 19 & 323665 & 7.984\\
ntu & 12 & 236901 & 5.570\\
ncl & 159 & 305025 & 7.180\\
oclc & 54 & 78792 & 1.848\\
queen & 33 & 311017 & 7.304\\
stowe & 58 & 30442 & 0.725\\
sydney & 156 & 769657 & 18.871\\
\cline{1-4}
\end{tabular}
\caption{\sf Sampling Overhead and Storage Requirement}
\label{table:cost}
\end{table*}

\begin{figure}
\centerline{\psfig{figure=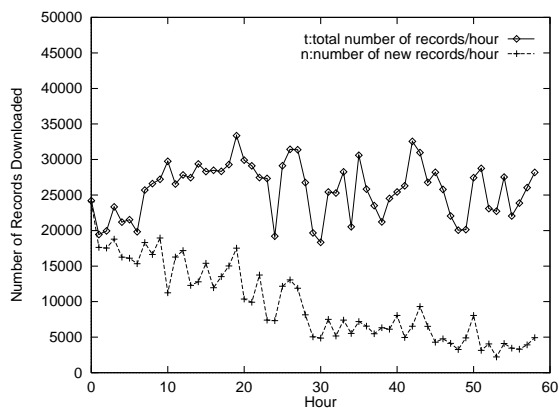}}
\caption{\sf Number of New Records Sampled Versus Total Number of Records Sampled}
\label{fig:t_boston}
\end{figure}

\begin{figure}
\centerline{\psfig{figure=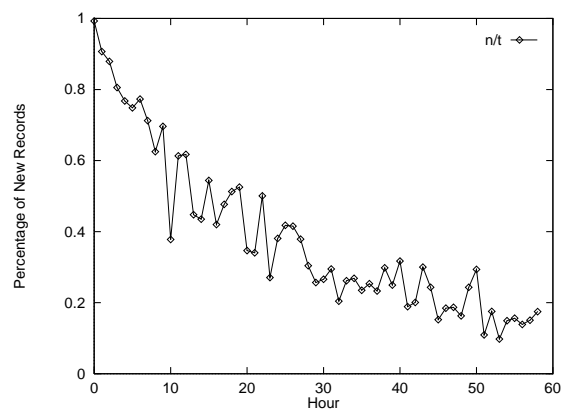}}
\caption{\sf Average Percentage of New Records Sampled} 
\label{fig:t_boston_p}
\end{figure}

Figure~\ref{fig:t_boston} depicts the number of new records sampled 
versus the total number of records sampled over time when
content sampling was performed on the Boston University Library.  
The figure indicates that large number of new records can be sampled at the
first hour of sampling.  
It is interesting to note that the number of new records
decreases as sampling continues.  This is further substantiated by
Figure~\ref{fig:t_boston_p} which shows the average percentage
of new records sampled per hour decreases over time.
Although the figures are derived from sampling the Boston University
Library, similar observation can be made for other Z39.50 databases.
The diminishing return effects suggest that early queries play
a more important role in probing the database content.
By examining the rate of return, indicated by the average percentage
of new records sampled per hour, one could determine if the
database content has been adequately sampled.

\subsection{Storage Requirement}

For each Z39.50 database, ZBroker maintains the content knowledge
consisting of tuple frequencies for the attributes author, title and subject. 
ZBroker also maintains global term dictionaries and a record id database 
for each Z39.50 database.
As shown in Table~\ref{table:totalcost}, a total of
137MBytes of storage space has been used to maintain the
eleven record id databases (one for each Z39.50 databases).
The size of record id database directly depends on the size of
the Z39.50 database concerned.

It takes only 2.8MBytes to store the tuple frequencies in
the content knowledge for each database.
The global term dictionaries also do not occupy too much
storage space and they only scale up sub-linearly due 
common terms appearing in different Z39.50 databases.

\begin{table}
\footnotesize
\centering
\begin{tabular}{|l|r|}
\cline{1-2}
Storage Components  & Storage Requirement(Mbytes)\\
\cline{1-2}
Global Term Dictionary(Title) & 48.2 (926254 terms) \\
Global Term Dictionary(Author) & 16.7 (320549 terms)\\
Global Term Dictionary(Subject) & 8.2 (158468 terms)\\
Content Knowledge/Database & 2.8 \\
All Record Id Databases  & 137.9\\
\cline{1-2}
\end{tabular}
\caption{\sf Storage Requirement}
\label{table:totalcost}
\end{table}

%----------------------------------------------------------------------

\section {9. CONCLUSIONS AND FUTURE WORKS}

In this paper, we describe a query routing system for Z39.50
databases that provide only Z39.50 query interface to their
content.  This system, known as ZBroker, demonstrates the viability
of using sampling technique to obtain content knowledge about
Z39.50 databases.
We have outlined the architecture, design and implementation of
ZBroker.  ZBroker is designed to cope with
changes in the database content.  It also consists of a 
web-based user interface that supports easy query formulation
and is able to forward queries to multiple selected Z39.50 databases.   

During the implementation of ZBroker, we encountered several
difficulties due to heterogeneous Z39.50 server
implementation. 
Although Z39.50 requires compliance from libraries
created by different vendors, there exists a wide variation 
of Z39.50 implementations among different	
libraries.  
For example, ``CD-ROM'' might be interpreted as
``CD ROM'' by some servers, but as a single term by others. 
This inconsistency will introduce some inaccuracy into the 
query routing results.

\subsection{Future Works}

As part of the future works, we plan to improve ZBroker
in a number of ways:

\begin{itemize}
  \item Note that ZBroker is not designed to be the only solution
	for routing queries to Z39.50 databases.  For Z39.50 databases
	that can export their full content information for query routing,
	it may be more appropriate to adopt the STARTS approach
	of ranking databases.  We plan to look into
	how a hybrid query routing system incorporating
	both STARTS and content sampling techniques can be developed.

  \item We plan to conduct more detailed performance evaluation
	experiments for ZBroker in order to determine the
	appropriate strategies to reduce the amount of time
	spent in content sampling and to improve upon the database 
	ranking techniques.

\end{itemize}

%-----------------------------------------------------
% Reference
%-----------------------------------------------------
\nocite{salton.b:88, cori:95, probability:84, metadata:97, sigir:98,
selecting:97}

\bibliography{dl99}

\begin{thebibliography}{10}

\bibitem{metadata:97}
M.~Baldonado, C.C.K. Chang, and L.~Gravano.
\newblock {M}etadata for {D}igital {L}ibraries: {A}rchitecture and {D}esign
  {R}ationale.
\newblock In {\em Proceedings of the 2nd ACM International Conference on
  Digital Libraries (DL'97)}, Pittsburgh, Pennsylvania, USA, 1997.

\bibitem{selecting:97}
M.~Buckland and C.~Plaunt.
\newblock {S}electing {L}ibraries, {S}electing {D}ocuments, {S}electing data.
\newblock In {\em Proceedings of ACM/ISDL}, 1997.

\bibitem{cori:95}
J.P. Callan, Z.~Lu, and W.B. Croft.
\newblock {S}earching {D}istributed {C}ollections {W}ith {I}nference
  {N}etworks.
\newblock In {\em Proceedings of the 18th Annual International ACM SIGIR
  Conference on Research and Development in Information Retrieval}, pages
  21--28, 1995.

\bibitem{sigir:98}
J.C. French, A.L. Powell, C.L. Viles, T.~Emmitt, and K.J. Prey.
\newblock {E}valuating {D}atabase {S}election {T}echniques: {A} {T}estbed and
  {E}xperiment.
\newblock In {\em Proceedings of the 21st ACM-SIGIR International Conference on
  Research and Development in Information Retrieval}, Melbourne, Australia,
  August 1998.

\bibitem{probability:84}
M.A. Golberg.
\newblock {\em An Introduction to Probability Theory with Statistical
  Applications}.
\newblock Plenum Press. New York and London, 1984.

\bibitem{starts:97}
L.~Gravano, Chen-Chuan~K. Chang, and H.~Garcia-Molina.
\newblock {STARTS}: {S}tanford {P}roposal for {I}nternet {M}eta-{S}earching.
\newblock In {\em Proceedings of the ACM SIGMOD Conference}, pages 207--218,
  Tucson, Arizona, USA, May 1997.

\bibitem{gloss:94}
L.~Gravano, H.~Garcia-Molina, and A.~Tomasic.
\newblock {T}he {E}ffectiveness of {G}l{OSS} for the {T}ext {D}atabase
  {D}iscovery {P}roblem.
\newblock In {\em Proceedings of the ACM SIGMOD International Conference on
  Management of Data}, pages 126--137, Minneapolis, Minnesota, May 1994.

\bibitem{gloss:99}
L.~Gravano, H.~Garcia-Molina, and A.~Tomasic.
\newblock {\em GlOSS}: {T}ext-{S}ource {D}iscovery over the {I}nternet.
\newblock {\em ACM Transactions on Database Systems(To Appear)}, 1999.

\bibitem{z3950}
National Information~Standard Organization(NISO).
\newblock {ANSI Z39.50: Information Retrieval Service and Protocol}, 1992.

\bibitem{fox:98}
J.~Powell and E.A. Fox.
\newblock Multilingual federated searching across heterogeneous collections.
\newblock {\em D-Lib Magazine}, September 1998.

\bibitem{salton.b:88}
G.~Salton.
\newblock {\em Automatic Text Processing, The Transformation, Analysis, and
  Retrieval of Information by Computer}.
\newblock Addison-Wesley Publishing Company, 1988.

\bibitem{broker.stru:96}
A.~Tomasic, L.~Gravano, C.~Lue, P.~Schwarz, and L.~Haas.
\newblock {D}ata {S}tructures for {E}fficient {B}roker {I}mplementation.
\newblock {\em ACM Transactions on Information Systems}, 15(3), July 1997.

\bibitem{voorhees:95}
G.~Towell, E.M. Voorhees, N.K. Gupta, and B.~Johnson-Laird.
\newblock {L}earning {C}ollection {F}usion {S}trategies for {I}nformation
  {R}etrieval.
\newblock In {\em Proceedings of the 12th Annual Machine Learning Conference},
  Lake Tahoe, July 1995.

\bibitem{dik:97}
B.~Yuwono and D.L. Lee.
\newblock {S}erver {R}anking for {D}istributed {T}ext {R}etrieval {S}ystems on
  the {I}nternet.
\newblock In {\em Proceedings of the 5th International Conference on Database
  Systems for Advanced Applications (DASFAA '97)}, pages 41--49, Melbourne,
  Australia, April 1997.

\end{thebibliography}

%-----------------------------------------------------
\end{document}